\documentclass[12pt]{article}

\begin{document}

\title{Is it possible to relate MOND with Ho\v{r}ava Gravity? }

\author{ Juan M. Romero\thanks{jromero@correo.cua.uam.mx},
R. Bernal-Jaquez\thanks{rbernal@correo.cua.uam.mx},
O. Gonz\'alez-Gaxiola\thanks{ogonzalez@correo.cua.uam.mx},
\\[0.5cm]
\it Departamento de Matem\'aticas Aplicadas y Sistemas,\\
\it Universidad Aut\'onoma Metropolitana-Cuajimalpa\\
\it M\'exico, D.F  01120, M\'exico\\[0.3cm]} 

\date{}

\pagestyle{plain}

\maketitle

\begin{abstract}
In this work we present a scalar field theory invariant under space-time anisotropic transformations with a dynamic exponent $z$. It is shown that this theory possess symmetries similar to Ho\v{r}ava gravity and that in the limit $z=0$ the equations of motion of the non-relativistic MOND theory are obtained.
This result  allow us to conjecture the existence of a Ho\v{r}ava type gravity that in the limit $z=0$ is consistent with MOND.\\  

Keyswords: Gravity; Mond; Horava Gravity   
\end{abstract}


\section{Introduction}

In recent years, various modifications have been proposed to general relativity.
One of the most notorious at a phenomenological level, is the relativistic version of the so called Modified Newtonian Dynamics (MOND) \cite{milgrom1:gnus}. Notoriously, without making use of dark matter, MOND successfully explains the anomalous dynamics of different astrophysical objects. For example,  it explains the rotation curves of different galaxies and the  Tully-Fisher  relation \cite{mond-feno:gnus}. 
Besides, MOND has an explanation for the so called Pioneer anomaly \cite{pioneer:gnus}.  We also have to mention that, MOND presents problems to predict the cluster's of galaxies dynamics, in this case however, the baryonic mass has not been measured  with certainty \cite{n-bekenstein:gnus}.
\\

MOND's  starting point is to assume that for small accelerations,  about $a_{0}\approx  10^{-8}cm/s^{2}$, Newton's second law takes on the form  \cite{milgrom1:gnus}
\begin{eqnarray}
\mu\left(\frac{|\vec a|}{a_{0}}\right)m  \vec a =\vec F,
\end{eqnarray}
with $\mu(u)$ defined as a function that satisfies 
\begin{eqnarray}
\mu(u)=\left\{
\begin{array}{ll}
1& \quad {\rm if} \quad u>> 1,\\
u&\quad {\rm if}\quad  u<< 1.
\end{array} \right.
\end{eqnarray}
The Newtonian regime is obtained if $u>>1$, meanwhile the MOND's regime is obtained if $u<<1$. In spite of is phenomenological success, MOND has problems at a theoretical level, for example, the energy for one particle is not conserved although it is conserved in several modified versions of the theory   \cite{yo:gnus}. \\

In the non-relativistic gravitational field regime, MOND is consistent with  \cite{milgrom2:gnus}
\begin{eqnarray}
\vec\nabla  \cdot\left( \mu\left( \frac{|\vec \nabla \phi|}{a_{0}} \right) \vec\nabla \phi \right)  =  4\pi G \rho,
\label{eq:aqual-0}
\end{eqnarray}
that in the MOND limit takes on the form 
\begin{eqnarray}
\vec\nabla  \cdot\left( \frac{|\vec \nabla \phi|}{a_{0}} \vec\nabla \phi \right)  =4\pi G\rho.
\label{eq:aqual-1}
\end{eqnarray}
This model gives a non-relativistic gravitational field description and does not present problems with conserved quantities.
However, using this model it is not possible to attack relativistic problems, such as  gravitational lenses or cosmological problems. To analyse these problems is necessary a relativistic MOND. Though there are several relativistic versions of MOND theory, the most complete is the so called  $TeVeS$ theory \cite{bekenstein:gnus}, see also \cite{moffat:gnus}. This version is compatible with Eq. (\ref{eq:aqual-1}) in the non-relativistic limit and  has been  successful in explaining several phenomenological facts.  It is worthy to mention that, with the present WMAP measurements, this theory cannot be discarded \cite{scordis:gnus,scordis-1:gnus}. With finer WMAP experimental measurements, this theory may be confirmed, modify or even discarded. \\

However, at a theoretical level, $TeVeS$ still has some inconsistencies,  for example,  it has some dynamical problems  as instabilities may appear \cite{inestable:gnus,giannios:gnus}. Therefore cannot be considered as a finished theory. In fact, modifications to this theory has been recently proposed \cite{inestable:gnus}  and also a new relativistic version of MOND has been recently proposed \cite{milgrom-nuevo:gnus}.\\

In the other hand, Ho\v{r}ava has recently proposed a modified version of  general relativity that in principle is renormalizable and free of ghosts  \cite{Horava:gnus}. This gravity assumes that space-time is compatible with the  anisotropic transformations 
\begin{eqnarray}
t\to b^{z}t,\qquad \vec x\to b\vec x,
\label{eq:escala-0}
\end{eqnarray}
where $z$  is a dynamic exponent. As a consequence of the transformation given in Eq. (\ref{eq:escala-0}),  the usual dispersion relation is substituted by 
\begin{eqnarray}
P^{2}_{0}-\tilde G\left( \vec P^{2}\right)^{z} =0,\qquad \tilde G={\rm constant}.
\label{eq:dispersion}
\end{eqnarray}
A remarkable point about this  dispersion relation is that, it is not obtained from a geodesic equation \cite{geo:gnus,geo-1:gnus,geo-2:gnus,geo-3:gnus}. Ho\v{r}ava gravity is not compatible with  Lorentz's  transformations neither  invariant under all the diffeomorphisms, nevertheless for long distances the usual relativity theory is regained.   This theory has some dynamical problems \cite{henneaux:gnus,henneaux-1:gnus,henneaux-2:gnus}, and cannot be considered as  complete. In fact, some recent proposals have been made in order to improve it \cite{blas:gnus}.\\

In this work, we will show a scalar field model compatible with the transformations stated in  Eq. (\ref{eq:escala-0}) whose dynamics in the limit  $z=0$ reduces to MOND Eq. (\ref{eq:aqual-1}). It is shown that this theory possess Weyl's symmetries similar to Ho\v{r}ava gravity. This allow us to conjecture that an anisotropic gravity  theory in space-time  could exist  such that, at $z=0$ reduces to a MOND type gravity; at $z=1$
standard gravity is regained and at $z=3$ we regain a Ho\v{r}ava type gravity compatible with quantum mechanics.  It worthy to mention that, the anisotropic transformations in Eq. (\ref{eq:escala-0}) are of  importance for the  $AdS/CFT$ non-relativistic duality by means of which efforts are made in order to relate condensed matter phenomena with string theory \cite{ads:gnus,ads-1:gnus}. This would make possible the existence of $AdS/CFT$ duality with a MOND type theory. \\

This manuscript is organized as follows: In section 2 we present our system and its equations of motion are studied. Section 3 is devoted to study the conserved quantities the system, in Section 4 we study the algebra of the conserved quantities and in Section 5 we present a  summary of our results.

\section{Action} 
Consider the Action invariant under the transformation Eq. (\ref{eq:escala-0})
\begin{eqnarray}
S&=&\int d x^{d}dt \left[a \left(\frac{\partial \phi( \vec x, t) }{\partial t}\right)^{\frac{z+d}{z}} +\gamma 
\left(\frac{\partial \phi( \vec x, t) }{\partial x^{i}} \frac{\partial \phi( \vec x, t) }{\partial x^{i}}  \right)^{\frac{z+d}{2}}
\right]\nonumber\\
&=&\int d x^{d}dt \left[a \left(\frac{\partial \phi( \vec x, t) }{\partial t}\right)^{\frac{z+d}{z}} +\gamma 
\left(\vec \nabla\phi\cdot \vec\nabla \phi  \right)^{\frac{z+d}{2}}
\right] .
\label{eq:accion}
\end{eqnarray}
Note that, if we define the non-zero elements of the metric $g_{\mu\nu}$ as  $g_{00}=1,g_{ij}=\delta_{ij}$, we have  $g={\rm det}g_{\mu\nu}$ and therefore the Action Eq. (\ref{eq:accion}) can be written as 

\begin{eqnarray}
S=\int d x^{d}dt \sqrt {g} \left[a \left(g^{00}\frac{ \partial \phi(\vec x, t) }{\partial t}    
\frac{\partial \phi( \vec x, t)}{\partial t} \right)^{\frac{z+d}{2z}} +\gamma
\left(g^{ij}   \frac{\partial \phi( \vec x, t) }{\partial x^{i}} \frac{\partial \phi( \vec  x, t) }{\partial x^{j}} \right)^{\frac{z+d}{2}}  \right],
\end{eqnarray}
this expression is invariant under Weyl's anisotropic transformations
\begin{eqnarray}
g_{00}\to \Omega^{2z}(\vec x,t) g_{00}, \qquad g_{ij}\to g_{ij} \Omega^{2}(\vec x,t). \label{eq:weyl}
\end{eqnarray}
This kind of symmetry is similar to the one present at Ho\v{r}ava gravity  \cite{Horava:gnus}.\\
The equation of motion for Eq. (\ref{eq:accion}) is

\begin{eqnarray}
a\left(\frac{z+d}{z} \right) \frac{\partial }{\partial t} \left( \frac{\partial \phi }{\partial t}\right)^{\frac{d}{z}}+
\gamma \left( z+d\right)
\frac{\partial }{\partial x_{i}} \left( \left(\frac{\partial \phi}{\partial x^{j}} \frac{\partial  }{\partial x^{j}} \right)^{\frac{z+d-2}{2}} \frac{\partial \phi }{\partial x^{i}}\right)=0,\nonumber
\end{eqnarray}
that can be written as
\begin{eqnarray}
a\left(\frac{z+d}{z} \right) \frac{\partial }{\partial t} \left( \frac{\partial \phi }{\partial t}\right)^{\frac{d}{z}}+
\gamma\left( z+d\right)
\vec \nabla \cdot \left( |\vec \nabla \phi|^{d+z-2} \vec \nabla \phi\right)=0.
\end{eqnarray}
If a source  $\rho$  is consider, the Action Eq. (\ref{eq:accion}) is now given by
\begin{eqnarray}
S=\int d x^{d}dt \left[a \left(\frac{\partial \phi}{\partial t}\right)^{\frac{z+d}{z}} +\gamma
\left(\vec \nabla \phi \cdot \vec \nabla \phi \right)^{\frac{z+d}{2}} +\phi \rho \right].
\label{eq:fuente}
\end{eqnarray}
If  under scaling the source is transformed as  $\rho \to \Omega^{-(z+d)}\rho,$  then Eq. (\ref{eq:fuente}) is invariant under Weyl's symmetry, Eq. (\ref{eq:weyl}). The  equation of motion  given by the Action Eq. (\ref{eq:fuente}) is
\begin{eqnarray}
a\left(\frac{z+d}{z} \right) \frac{\partial }{\partial t} \left( \frac{\partial \phi }
{\partial t}\right)^{\frac{d}{z}}+\gamma \left(z+d\right)
\vec \nabla \cdot \left( |\vec \nabla \phi|^{d+z-2} \vec \nabla \phi\right)=-\rho.
\end{eqnarray}
It can be noticed that, at the limit $z \to 0$ the first term of the Action in  Eq. (\ref{eq:fuente}) is constant and we can obtain the effective Action as
\begin{eqnarray}
S=\int d x^{d} \left[ \gamma
\left(\vec \nabla \phi \cdot \vec \nabla \phi \right)^{\frac{d}{2}} +\phi \rho \right], 
\end{eqnarray}
whose equation of motion is
\begin{eqnarray}
\gamma d\vec \nabla \cdot \left( |\vec \nabla \phi|^{d-2} \vec \nabla \phi\right)=-\rho.
\end{eqnarray}
If $d=3$  and $\gamma=-1/(12\pi Ga_{0})$ we obtain the MOND non-relativistic equation of motion Eq. (\ref{eq:aqual-1}), \cite{milgrom2:gnus}. \\

Therefore the system under study contains the non-relativistic MOND's theory and coincides with Ho\v{r}ava gravity symmetries. This fact, make it possible to conjecture the existence of a Ho\v{r}ava type gravity that, in $z=0$ limit reduces to MOND. Note that, as the Ho\v{r}ava gravity must be valid in the quantum regime, the fundamental constant is Planck's mass $M_{P}$. This constant seems to be non-related with MOND's fundamental constant $a_{0}$. However, $a_{0}$ can be written as $a_{0}\approx m_{N}c\left(6M_{P}^{3}t_{p}\right)^{-1},$ where $m_{N}$ is the proton mass. It is possible that in this conjectured gravity this type of relations could arise in a natural way.  It can be noticed that, in this relation we have a collection of apparently dissimilar quantities, however, by means of the Holographic principle,  this type of relations appear  \cite{marugan:gnus}.

\section{Noether's theorem}

In this section, we will find out  the conserved quantities of the Action Eq. (\ref{eq:accion}). First, note that the canonical momentum is given by
\begin{eqnarray}
\Pi=\frac{\partial {\cal L}}{\partial \dot \phi}=a\frac{z+d}{z}\left(\dot \phi\right)^{\frac{d}{z}},
\end{eqnarray}
therefore, the equation of motion can be written as
\begin{eqnarray}
\frac{\partial \Pi}{\partial t} +\gamma (z+d)
\vec \nabla \cdot \left( |\vec \nabla \phi|^{d+z-2} \vec \nabla \phi\right)=-\rho.
\end{eqnarray}
Considering Noether's theorem, we know that the temporal part of 
\begin{eqnarray}
\int d^{d}J_{\mu}=\int dx^{d} \left( \frac{\partial {\cal L}}{\partial (\partial^{\mu} \phi)} 
\frac{\partial \phi}{\partial x^{\nu}}-g_{\mu\nu} {\cal L} \right)\delta x^{\nu}
\end{eqnarray}
is conserved.  Taking into account that the Action Eq. (\ref{eq:accion}) is invariant under temporal translations, we can conclude that the  Hamiltonian is conserved 
\begin{eqnarray}
H=\int dx^{d}{\cal H}= \int dx^{d} \left(\frac{ad}{z}\left(\frac{z}{a(z+d)}\right)^{\frac{d+z}{d}} 
\Pi^{\frac{z+d}{d}}-\gamma|\vec \nabla \phi|^{z+d}\right).
\end{eqnarray}
Besides, the  momentum is conserved
\begin{eqnarray}
P_{i}=\int dx^{d} p_{i}=\int dx^{d} \Pi\frac{\partial \phi}{\partial x^{i}}=
\int dx^{d} a\frac{z+d}{z}\left(\dot \phi\right)^{\frac{d}{z}} \frac{\partial \phi}{\partial x^{i}}
\end{eqnarray}
as well as the angular momentum
\begin{eqnarray}
L_{i}=-\int dx^{d} \Pi \epsilon_{ijk}x_{j}\frac{\partial \phi}{\partial x^{k}}=
-\int dx^{d}  a\frac{z+d}{z}\left(\dot \phi\right)^{\frac{d}{z}}
\epsilon_{ijk}x_{j}\frac{\partial \phi}{\partial x^{k}}. 
\end{eqnarray}
The scaling generator  
\begin{eqnarray}
D=\int dx^{d}\left( zt{\cal H}+p_{i}x^{i}\right)
\end{eqnarray}
is also conserved.
This quantities form the algebra
\begin{eqnarray}
\{H,P_{i}\}&=&0,\\
\{H,L_{kl}\}&=&0,\\
\{H,D\}&=&zH,  \\
\{D,P_{i}\}&=&-P_{i} , \\
\{D,L_{i}\}&=&0, \\
\{P_{i},P_{j}\}&=&0, \\
\{P_{i},L_{j}\}&=&\epsilon_{ijm}P_{m}, \\
\{L_{i},L_{j}\}&=&\epsilon_{ijk}L_{k}  .
\end{eqnarray}
This type of algebraic relations are characteristic of anisotropic scale invariant systems with dynamic exponent $z$.

\section{Summary}
In this work we have presented a scalar field theory invariant under space-time anisotropic transformations with a dynamic exponent $z$. It is shown that this theory possess symmetries similar to Ho\v{r}ava gravity, in particular Weyl's symmetries. Also, it is shown that in the limit $z=0$ the equations of motion of the non-relativistic MOND theory are obtained. This result make it possible to conjecture the existence of a Ho\v{r}ava type gravity that in the limit $z=0$ is consistent with MOND. Also, conserved quantities and their algebraic relations have been studied.



\end{document}